\documentclass[10pt, conference, letterpaper]{IEEEtran}

\newcommand*{\TechReport}{}%

\ifdefined\TechReport
\else
\usepackage{enumitem}
\fi
\usepackage[pass]{geometry}
\usepackage{url,xspace,graphicx}
\usepackage{subcaption}
\usepackage{units}
\usepackage{float}
\usepackage{listings,mdframed}
\usepackage{array}
\usepackage{verbatim}
\usepackage{color}
\ifdefined\TechReport
\usepackage[hidelinks]{hyperref}
\fi

\newlength{\grafflecm}
\usepackage{pslatex}
\usepackage{clrscode}
\usepackage{theorem}
\theoremheaderfont{\bfseries}
\theoremstyle{plain}
\theorembodyfont{\rmfamily}

\usepackage{amsmath}
\usepackage{amsfonts}
\usepackage{amssymb}

\IEEEoverridecommandlockouts

\begin{document}

\clubpenalty=10000 
\widowpenalty = 10000

\interfootnotelinepenalty=10000

\title{
\ifdefined\TechReport
\hspace{3mm}
\fi
The Case for Data Plane Timestamping in SDN
\ifdefined\TechReport
\newline
\newline
				\large Technical Report\textsuperscript{\ensuremath\diamond}\thanks{\textsuperscript{\ensuremath\diamond}This technical report is an extended version of~\cite{swfanDPT}, which was accepted to the IEEE INFOCOM Workshop on Software-Driven Flexible and Agile Networking (SWFAN) 2016.}, February 2016
\fi
}

\author{
{Tal Mizrahi, Yoram Moses\textsuperscript{\ensuremath*}\thanks{\textsuperscript{\ensuremath*}\scriptsize Yoram Moses is the Israel Pollak academic chair at Technion.}}\\
Technion --- Israel Institute of Technology\\
Email: \{dew@tx, moses@ee\}.technion.ac.il
}

\maketitle

\ifdefined\CutSpace\vspace{-1mm}\fi
\begin{abstract}
This paper presents the case for Data Plane Timestamping (DPT). We argue that in the unique environment of Software-Defined Networks (SDN), attaching a timestamp to the header of all packets is a powerful feature that can be leveraged by various diverse SDN applications. We analyze three key use cases that demonstrate the advantages of using DPT, and show that SDN applications can benefit even from using as little as one bit for the timestamp field.
\end{abstract}


\ifdefined\CutSpace\vspace{-1mm}\fi
\section{Introduction}
\ifdefined\CutSpace\vspace{-1mm}\fi
\label{IntroSec}
\subsection{DPT in a Nutshell}
Time and synchronized clocks are used in network protocols for various purposes, such as network telemetry~\cite{Y1731,INT,kimband}, Time-Sensitive Networking (TSN)~\cite{IEEETSN}, and time-triggered network updates~\cite{OpenFlow1.5}. What if we had a \textbf{timestamp} attached to \textbf{every} packet in the network?

In this paper we make the case for Data Plane Timestamping (DPT). We argue that adding a timestamp to the header of \textbf{every packet} is a powerful tool that is useful for various diverse applications. 

\begin{sloppypar}
DPT is especially relevant in Software-Defined Networks (SDN) for two main reasons: (i) An SDN is a locally-administered environment, where a DPT header can be inserted by \textbf{ingress} switches and removed by \textbf{egress} switches, and (ii) The current trend in SDN~\cite{bosshart2014p4,P4spec,ONFPIF} assumes protocol-independent forwarding, providing the flexibility to add and remove any header, and to take (match) decisions based on any header field.
\end{sloppypar}

The \textbf{Data Plane Timestamp Header (DPTH)} indicates the time at which the packet enters the network. This labeling method is both flexible and uniform; the generic label can be flexibly used by multiple different SDN applications at the same time, and it allows switches to treat each packet consistently, based on its timestamp value.

This paper focuses on three main use cases that demonstrate the merit of DPT:
\ifdefined\TechReport
\begin{itemize}
\else
\begin{itemize}[leftmargin=*,topsep=0pt,itemsep=-1ex,partopsep=1ex,parsep=1ex]
\fi
	\item \textbf{Network telemetry.} DPT allows to measure and monitor the network delay and packet loss using the timestamp, without the need for any additional metadata to be exchanged between switches.
	\item \textbf{Consistent network updates.} The timestamp field can play the role of a configuration \emph{version tag}~\cite{reitblatt2012abstractions}, thereby reducing the management overhead of consistent updates.
	\item \textbf{Load balancing.} The DPTH allows the use of time-division for forwarding elephant flows over multiple paths, allowing higher throughput than other stateless load balancing methods.
\end{itemize}
\vspace{2mm}

DPT allows switches to take packet processing decisions based on the timestamp field. Since match procedures in SDN switches allow header fields to be partially masked~\cite{OpenFlow1.5,P4spec}, we leverage the work of~\cite{Infocom-TimeFlip} to define DPT-based \textbf{time ranges}. Thus, policies or paths can be restricted to specific timestamp ranges.



DPT raises an inevitable question: is it practical to add a new header to \textbf{all packets} in the network? Many of the networks in which SDN is deployed or considered use network overlay protocols, e.g., VXLAN~\cite{vxlan}, Geneve~\cite{geneve}, and NSH~\cite{nsh}. These overlay protocols provide inherent extensibility for adding metadata fields, and are therefore DPT-friendly. 
Furthermore, as we discuss in this paper, a compact timestamp, sometimes even a single-bit, may be sufficient in some applications; since an SDN is a locally-controlled environment, a single-bit DPTH can be accommodated by an unused field in the packet header.

\subsection{Contributions} 
The main contributions of this paper are:
\ifdefined\TechReport
\begin{itemize}
\else
\begin{itemize}[leftmargin=*]
\fi
	\item We make the case for attaching a timestamp to \textbf{all packets} in SDNs.
	\item Three key use cases that can benefit significantly from using DPT are presented and analyzed.
	\item We analyze a simple case where a one-bit timestamp is attached to all packets, demonstrating the benefits that can be achieved using this small overhead.
	\item Using experimental evaluation we show the merits of DPT in each of the three use cases.
\end{itemize}

\ifdefined\TechReport
\else
Due to space limits, some of the detailed discussion is presented in a technical report~\cite{DPT-TR}.
\fi

\subsection{Related Work}
Packet timestamping has been proposed for various purposes, such as measurement~\cite{TCPhighPerf,Y1731,INT,mittal2015timely,nshtimestamp}, and clock synchronization~\cite{mills2010rfc,IEEE1588}. 
The current paper presents DPT, an approach that adds a timestamp to \textbf{all packets}, thereby allowing various SDN applications to use the timestamp for different purposes.

TCP supports an extension~\cite{TCPhighPerf} that adds a timestamp to all packets, allowing to compute the round-trip-time~\cite{TCPhighPerf}. The current paper proposes to attach a timestamp to \textbf{all} packets, including non-TCP traffic. Furthermore, our solution does not require the end points to be aware of the DPT, since it is used only within the premises of the SDN network.

\ifdefined\TechReport
\else
The work of Lamport~\cite{lamport1978time} suggested that distributed applications can use \emph{logical clocks}, which produce monotonically increasing \emph{sequence numbers} instead of \emph{timestamps}. In this paper we show that there are advantages to using a DPT that represents the time-of-day, that cannot be achieved by using sequence numbers.
\ifdefined\TechReport
These advantages are further discussed in~\ref{DiscussSec}.
\else
These advantages are further discussed in~\cite{DPT-TR}.
\fi
\fi



\begin{sloppypar}
The work of~\cite{TimedConsistent} suggested to use accurate time to schedule consistent updates, thereby reducing the update duration. In this paper we show that the use of \textbf{in-band timestamps} eliminates the need for version tags, and reduces the load on the SDN controller.
TimeFlip~\cite{Infocom-TimeFlip} introduced the use of time ranges in switches' TCAMs using an internally-generated timestamp. The current paper generalizes the scope of this work; we show that TimeFlips can be applied at a network-wide scale using a network-wide timestamp, and that DPT-based TimeFlips can be implemented using the ternary match logic of common open source SDN switches.
\end{sloppypar}

\section{An Overview of DPT}
\subsection{Timestamping Everything}
We propose to timestamp \textbf{all} packets. Every packet that enters the network's administrative domain is timestamped. Every packet that is forwarded through the network undergoes the following steps, corresponding to Fig.~\ref{fig:TimestampNetwork}:

\ifdefined\TechReport
\begin{enumerate}
\else
\begin{enumerate}[leftmargin=*,topsep=0pt,itemsep=-1ex,partopsep=1ex,parsep=1ex]
\fi
	\item The ingress switch attaches a DPTH to all packets. The ingress switch may either be a hardware switch or a virtual (software) switch, i.e., a vSwitch.
	\item Packets are forwarded through the network with the DPTH. Switches can use the DPTH in their match-action processing. In an environment that uses Virtual Network Functions (VNF), the DPTH can be used by VNFs as well.
	\item The egress switch removes the DPTH from the header of every packet.
\end{enumerate}

\begin{figure}[htbp]
	\centering
  \fbox{\includegraphics[width=.3\textwidth]{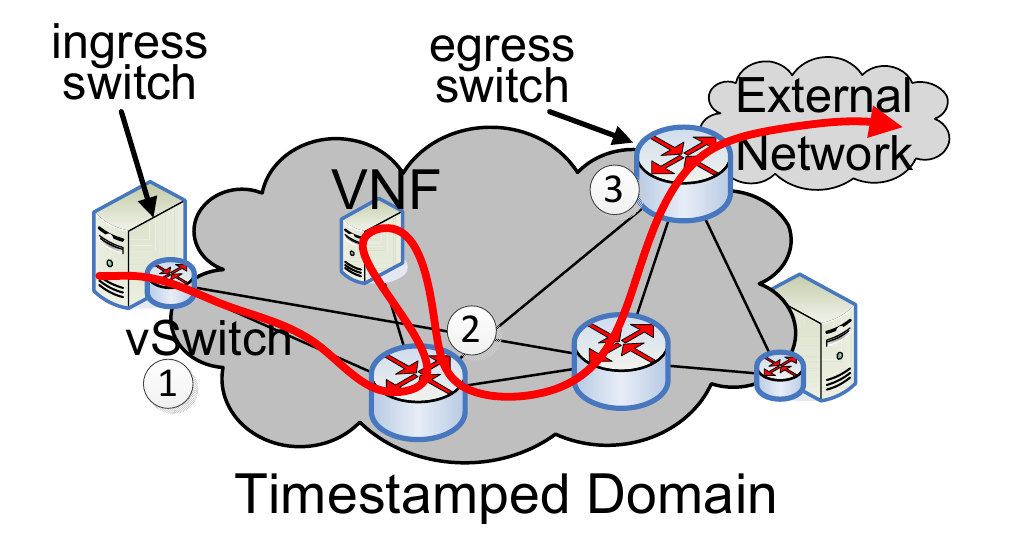}}
  \caption{Timestamping all packets.}
  \label{fig:TimestampNetwork}
\end{figure}

In this paper we focus on the use of DPT in SDN switches. It should be noted that environments that use Network Function Virtualization (NFV) can also benefit from DPT, as the timestamps can be similarly used in the processing of VNFs. 

\subsection{Using the Timestamp}
\label{UsingTimestampsSec}
Since a DPTH is integrated into every packet, it can be used in the switches' packet processing procedure. Two of the most promising on-going SDN efforts, P4~\cite{P4spec} and OF-PI~\cite{ONFPIF}, allow to flexibly parse and process packet headers.

Specifically the DPTH can be used in the switch match procedure; in addition to conventional match fields such as the 5-tuple, a flow match entry may also include the timestamp field. Since the switch's match process~\cite{P4spec,OpenFlow1.5} allows to define masks for each match field, it is possible to define time ranges, by masking part of the timestamp field (see Example 1 below). By defining a time range in a flow entry, we are confining the match rule to a specific range of times.

In the context of this paper we focus on two types of time ranges (as defined in~\cite{Infocom-TimeFlip}): (i) \textbf{extremal ranges}, i.e., ranges of the form $T \geq T_0$ (Fig.~\ref{fig:ExtremalRange}), and (ii) \textbf{periodic ranges}, defining a set of values with a periodic pattern (Fig.~\ref{fig:PeriodicRange}).

\begin{figure}[htbp]
	\centering
  \begin{subfigure}[t]{.23\textwidth}
  \centering
  \fbox{\includegraphics[height=1.2\grafflecm]{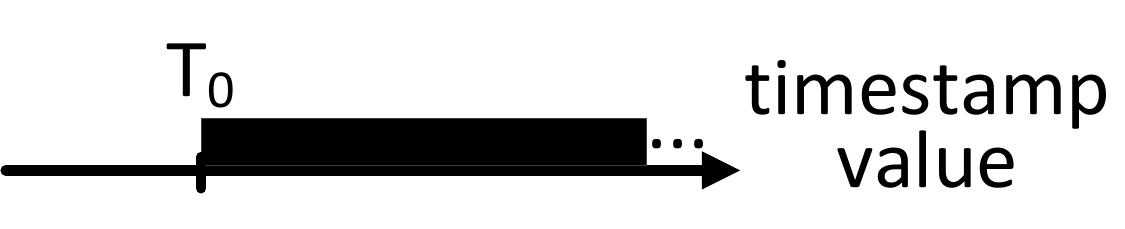}}
  \caption{Extremal range.}
  \label{fig:ExtremalRange}
  \end{subfigure}%
  \begin{subfigure}[t]{.25\textwidth}
  \centering
  \fbox{\includegraphics[height=1.2\grafflecm]{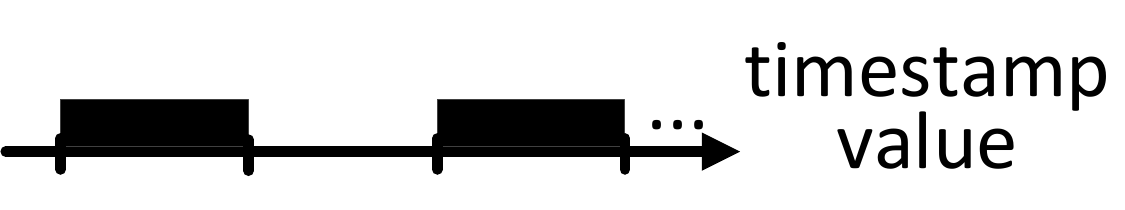}}
	\captionsetup{justification=centering}
  \caption{Periodic range.}
  \label{fig:PeriodicRange}
  \end{subfigure}%
  \caption{Time ranges.}
  \label{fig:Ranges}
\end{figure}

\subsection{Timestamp Format}
\label{FormatSec}
Various different timestamp formats are used in network protocols~\cite{mills2010rfc,IEEE1588,Y1731,rfc3339}. 
In this paper we choose to use the 64-bit NTP timestamp format~\cite{mills2010rfc}, although the DPT concept applies to other timestamp formats as well.\footnote{In fact, DPT can be used with one of the previously defined in-band timestamp encapsulations (e.g.,~\cite{TCPhighPerf,INT,nshtimestamp}), even though they were defined for different purposes.}
This time format represents the time elapsed since the base date, which is 1 January, 1900. The time format consists of two 32-bit fields: (i) Time.Sec: the integer part of the number of seconds since the base date, and (ii) Time.Frac: the fractional part of the number of seconds.

\ifdefined\TechReport
The size of the DPT timestamp is further discussed in~\ref{DiscussSec}.
\else
The size of the DPT timestamp is further discussed in~\cite{DPT-TR}.
\fi

\emph{Example 1.} Fig.~\ref{fig:RangeExamples} illustrates two time ranges, represented using the NTP time format. The `*' symbol represents bits that are masked.

\begin{figure}[htbp]
	\centering
  \begin{subfigure}[t]{.23\textwidth}
  \centering
  \fbox{\includegraphics[height=1.45\grafflecm]{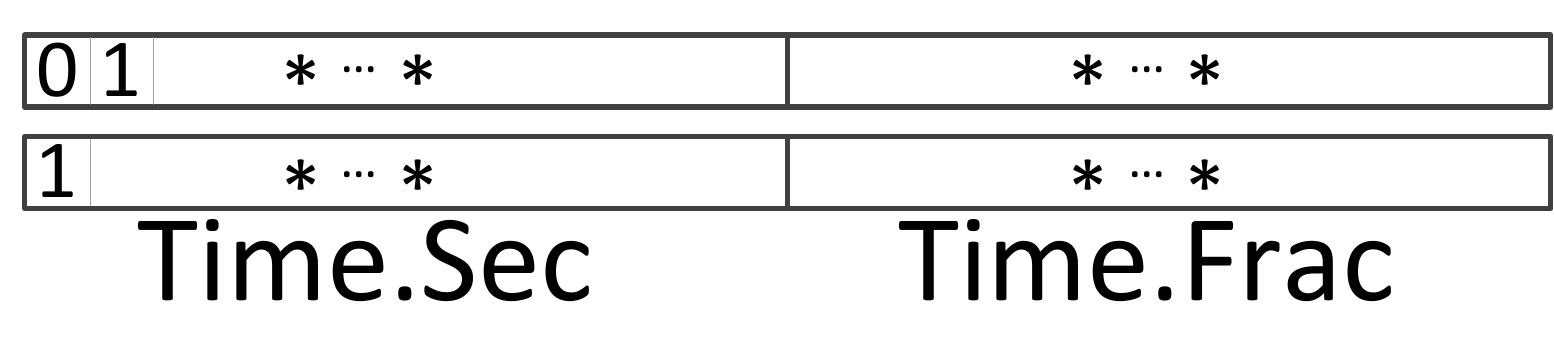}}
	\captionsetup{justification=centering}
  \caption{The extremal range $T \geq 2^{30}$ sec. Represented using the $31^{st}$ and $32^{nd}$~bits of~Time.Sec. This makes use of two~match~entries.}
  \label{fig:ExtremalExample}
  \end{subfigure}%
  \begin{subfigure}[t]{.25\textwidth}
  \centering
  \fbox{\includegraphics[height=1.45\grafflecm]{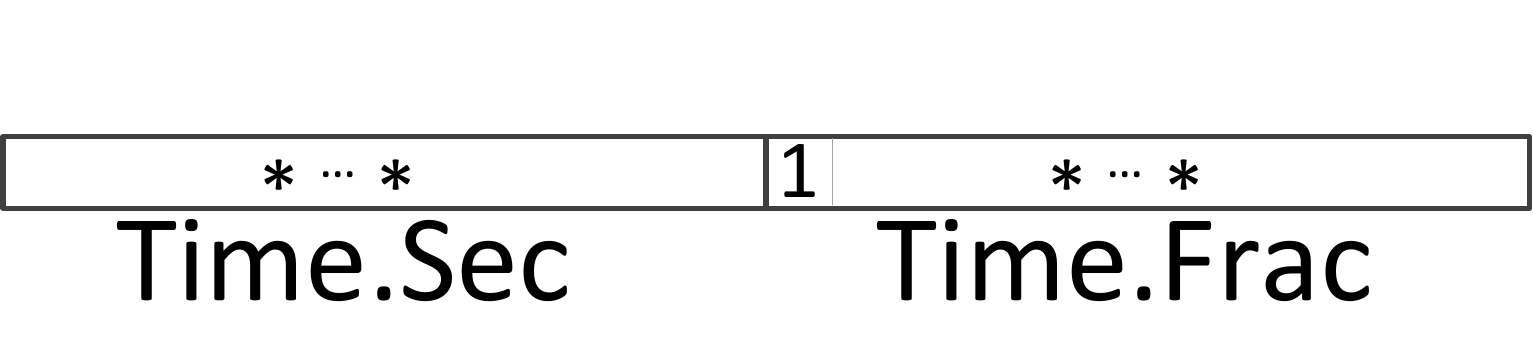}}
	\captionsetup{justification=centering}
  \caption{A periodic range that is active during the last half of every second. Uses the most significant bit of~Time.Frac.}
  \label{fig:PeriodicExample}
  \end{subfigure}%
	\ifdefined\CutSpace\vspace{-3mm}\fi
  \caption{Time range examples.}
  \label{fig:RangeExamples}
	\ifdefined\CutSpace\vspace{-3mm}\fi
\end{figure}

As shown in Fig.~\ref{fig:ExtremalExample}, representing a time range may require multiple match entries. As discussed in~\cite{Infocom-TimeFlip}, measures can be taken to reduce the number of entries. Specifically, in Sec.~\ref{OneBitSec} we show that when a \textbf{one-bit} timestamp is used, every time range requires \textbf{a single} match entry.

\section{DPT Use Cases}
An in-band timestamp can be used for various purposes. We focus on three applications that can greatly benefit from using the timestamp.

\subsection{Network Telemetry}
\label{TelemetrySec}
Performance measurement and monitoring is of key importance in large-scale networks, allowing to detect network faults, anomalies, and congestion, and to enforce a Service Level Agreement (SLA).

\ifdefined\TechReport
\subsubsection{Timestamp-based Measurement}

DPT can be used for accurate delay measurement. As illustrated in Fig.~\ref{fig:SimpleDM}, a timestamped packet sent from switch $S_1$ to switch $S_2$ allows $S_2$ to compute the one-way delay from $S_1$ to $S_2$, assuming that the two switches have synchronized clocks. The one-way delay in this case is given by $T_2 - T_1$.

\begin{figure}[htbp]
	\centering
  \fbox{\includegraphics[width=.2\textwidth]{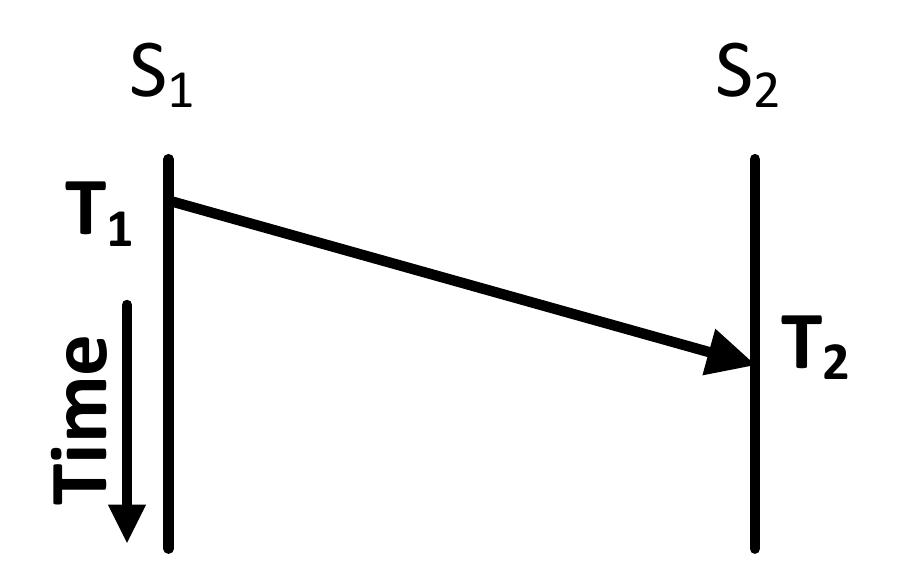}}
  \caption{Timestamp-based delay measurement.}
  \label{fig:SimpleDM}
\end{figure}

\subsubsection{Color-based Measurement}
The measurement method described above is useful for measuring network \textbf{delay}. In contrast, coloring-based passive measurement~\cite{coloring} allows for both \textbf{delay} and \textbf{loss} measurement.
\else
Various protocols (e.g.~\cite{Y1731,INT,kimband}) use timestamped packets to measure \textbf{delay}. We introduce a timestamp-based variant of coloring-based passive measurement~\cite{coloring}, which allows for both \textbf{delay} and \textbf{loss} measurement.
\fi

In the coloring approach of~\cite{coloring} the header of every packet sent between two measurement agents,  $S_1$ and $S_2$, includes a binary \emph{color} bit, either `0' or `1'. The color bit divides the traffic into consecutive blocks of packets, allowing $S_1$ and $S_2$ to process each block separately.

According to~\cite{coloring}, the color is toggled periodically (e.g., every minute), so that each color is used for a fixed time interval. In the context of SDN this approach would require the SDN controller to periodically update the configuration of $S_1$ each time the color needs to be toggled.

DPT alleviates the need for a color bit. The switches $S_1$ and $S_2$ can derive the color from the in-band timestamp\footnote{Note that the DPT is generated by the ingress switch, which may or may not be $S_1$.} value, by using periodic time ranges, as shown in Fig.~\ref{fig:ColorPeriodic}. 

\begin{figure}[htbp]
	\centering
  \fbox{\includegraphics[width=.3\textwidth]{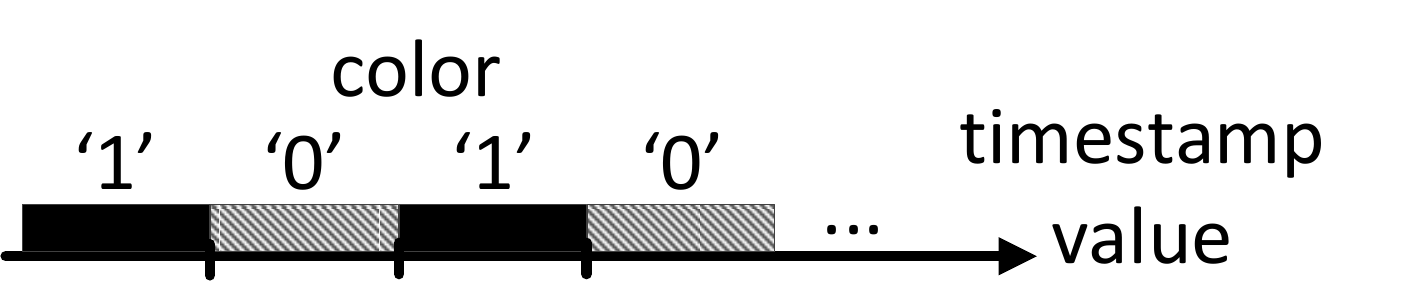}}
  \caption{Coloring based on timestamp ranges.}
  \label{fig:ColorPeriodic}
\end{figure}

\textbf{The advantage} of our approach is that the SDN controller does not need to periodically perform an update in $S_1$ that toggles the color bit; instead, the color is directly derived from the DPT. 

\ifdefined\TechReport
Loss and delay measurement are performed separately for each block.
\fi

\textbf{Loss measurement (LM).} Each of the two switches that take part in the measurement maintains two counters per flow. The sender $S_1$, maintains $CS0$ and $CS1$, one counter for each color, and the receiver $S_2$, maintains $CR0$ and $CR1$. 

\ifdefined\TechReport
\begin{figure}[htbp]
	\centering
  \begin{subfigure}[t]{.23\textwidth}
  \centering
  \fbox{\includegraphics[height=6.5\grafflecm]{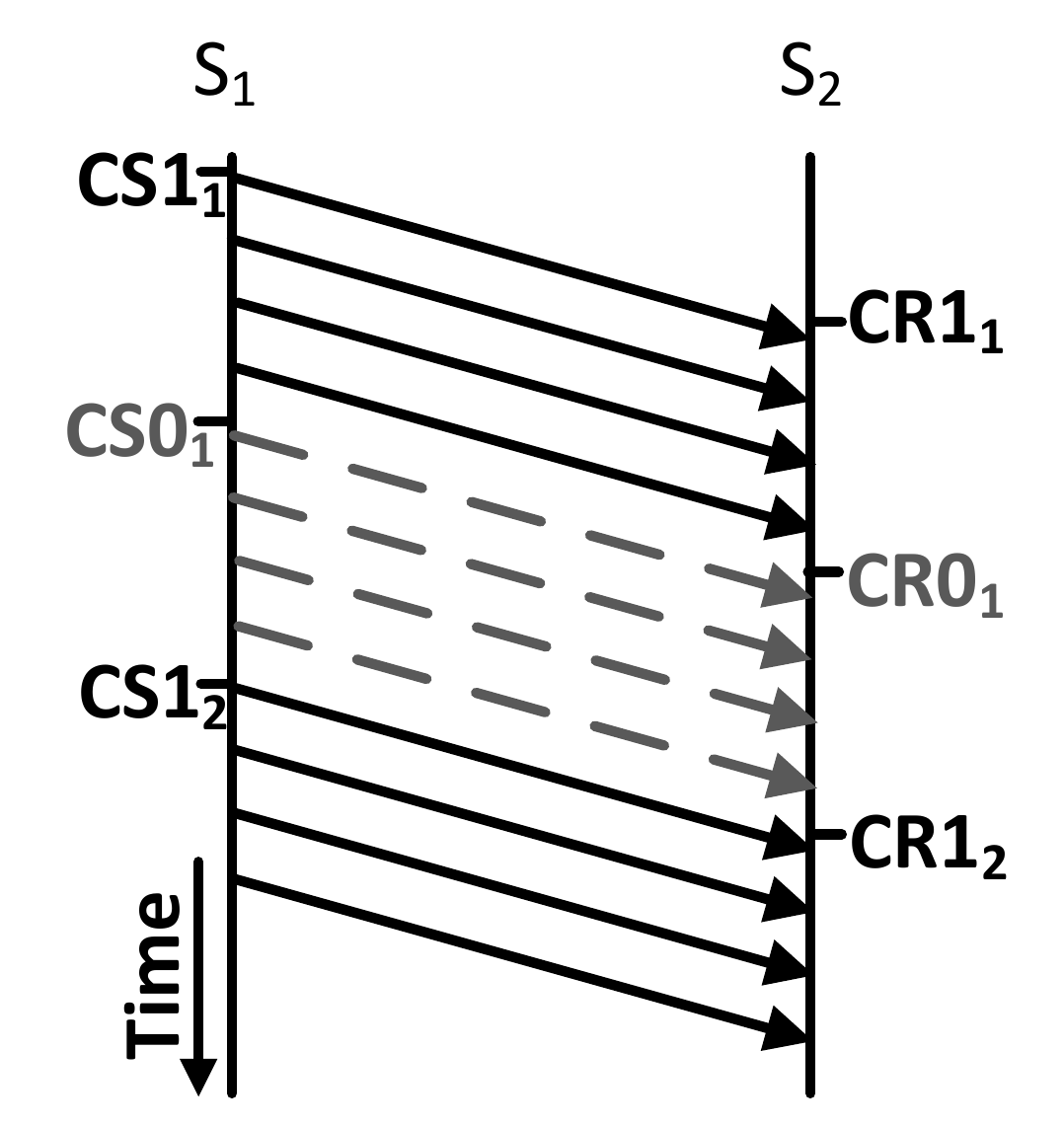}}
  \caption{Coloring-based loss measurement.}
  \label{fig:ColLM}
  \end{subfigure}%
  \begin{subfigure}[t]{.25\textwidth}
  \centering
  \fbox{\includegraphics[height=6.5\grafflecm]{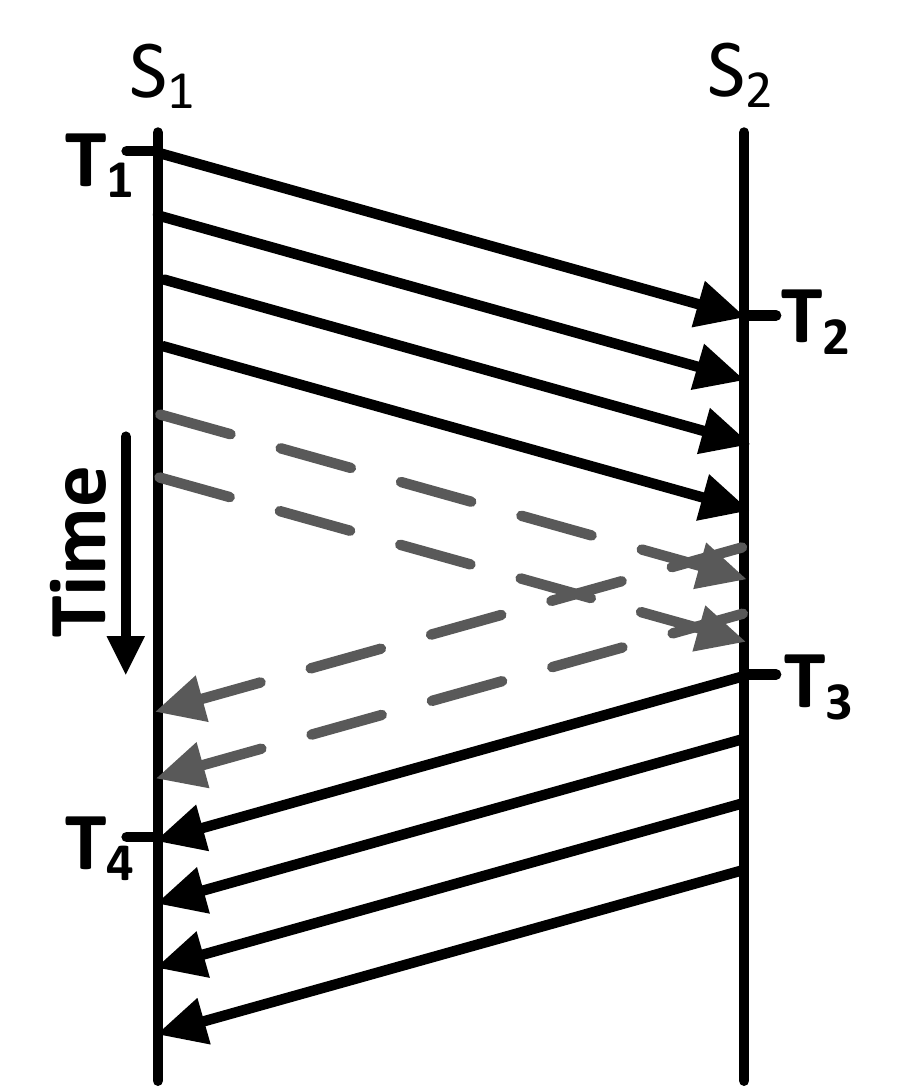}}
	\captionsetup{justification=centering}
  \caption{Coloring-based delay measurement.}
  \label{fig:ColDM}
  \end{subfigure}%
  \caption{Coloring-based method.}
  \label{fig:Color}
\end{figure}
\else
\begin{figure}[htbp]
	\centering
  \fbox{\includegraphics[width=.25\textwidth]{ColLMWide}}
  \caption{Coloring-based loss measurement.}
  \label{fig:ColLM}
\end{figure}
\fi

Fig.~\ref{fig:ColLM} depicts three consecutive blocks of traffic. At the end of each block the controller collects the counter values from each of the switches. Hence, after the second block has completed the controller can compute the $S_1$-to-$S_2$ packet loss during the first block:

\ifdefined\CutSpace\vspace{-3mm}\fi
\begin{equation}
(CS1_2 - CS1_1) - (CR1_2 - CR1_1)
\end{equation}

\ifdefined\TechReport
\textbf{Delay measurement (DM).} Each of the two switches uses the first packet of each block as a reference for delay measurement. Switch $S_1$ keeps the time of transmission of the first packet of the block, and $S_2$ keeps the time of reception of this packet. The controller periodically collects these timestamps, to be used for computing the delay.

The one-way delay from $S_1$ to $S_2$, assuming synchronization between the two switches is:
\begin{equation}
T_2 - T_1
\end{equation}

Based on four timestamps collected from two traffic blocks, as shown in Fig.~\ref{fig:ColDM}, the controller can compute the two-way delay between the two switches: 

\begin{equation}
\label{Eq:TwoWayDelay}
(T_4 - T_1) - (T_3 - T_2)
\end{equation}

Note that the computation of Eq.~\ref{Eq:TwoWayDelay} does not require the two ends to be synchronized.


One of the problems raised in~\cite{coloring} is that packets may arrive to $S_2$ out-of-order, and therefore the timestamp measured by $S_1$ may refer to a different packet than the one measured in $S_2$. DPT provides an inherent solution for this problem; when a switch measures a DM timestamp, it also keeps the DPTH value, extracted from the corresponding packet. The controller can then retrieve the measured timestamps and DPTH values from the two switches, and verify that the two readings have a matching DPTH value.
\else
\textbf{Delay measurement (DM).} Coloring can also be used for measuring delay, as further discussed in a technical report~\cite{DPT-TR}. For the sake of brevity, in the current paper we focus on loss measurement.
\fi

\subsection{Consistent Network Updates}
Consistent network updates have been widely analyzed in the literature. An update is consistent~\cite{reitblatt2012abstractions} if every packet is processed by all the switches along its path either according to the `old' configuration, before the update, or according to the `new' one. As a test case, we focus on the \emph{two-phase updates} of Reitblatt et al.~\cite{reitblatt2012abstractions}; all packets include a version tag, indicating for each packet whether it is processed according to the `old' configuration, or according to the `new' one. The version tag allows the update to be performed in two phases, as specified in Fig.~\ref{fig:TwoPhaseAlgo}. 

\begin{figure}[!h]
	\hrule
	\ifdefined\CutSpace\vspace{-3mm}\fi
  \begin{codebox}
    \Procname{$\proc{Two-phase Update}$}
		\li Controller sends \textbf{new} configuration to switches.
		\li Controller enables \textbf{new} version tag in ingress switches.
  \end{codebox}
	\ifdefined\CutSpace\vspace{-3mm}\fi
  \hrule
  \caption{Consistent two-phase update, as in~\cite{reitblatt2012abstractions}.}
  \label{fig:TwoPhaseAlgo}
	\ifdefined\CutSpace\vspace{-3mm}\fi
\end{figure}

The DPT approach provides an inherent version tag to all packets. The value of the timestamp progressively increases according to the local clocks of the ingress switches. For each update the controller can define a threshold value $T_{thr}$, such that the \textbf{new} configuration is effective during the time range $T \geq T_{thr}$.

\begin{figure}[!h]
	\hrule
	\ifdefined\CutSpace\vspace{-3mm}\fi
  \begin{codebox}
    \Procname{$\proc{Timestamp-based One-phase Update}$}
		\li Controller sends \textbf{new} configuration and $T_{thr}$ to switches. \\
		    No need to update tagging policy of ingress switches.
  \end{codebox}
	\ifdefined\CutSpace\vspace{-3mm}\fi
  \hrule
  \caption{Consistent timestamp-based \textbf{one-phase} update.}
  \label{fig:OnePhaseAlgo}
	\ifdefined\CutSpace\vspace{-3mm}\fi
\end{figure}

The timestamp allows `two-phase' updates to be performed using a single\footnote{The two-phase approach, as well as our DPT approach, often requires an additional phase for garbage collection, where the old configuration is removed from the switches. This is further discussed in Sec.~\ref{EvalSec}.} phase (Fig.~\ref{fig:OnePhaseAlgo}). The controller is no longer required to contact the ingress switches to initiate an update. The second phase is replaced by the DPT which automatically assigns timestamps. 

Another important advantage of our approach is that it significantly simplifies the management overhead of two-phase updates.  
For example, consider an SDN application that performs multiple updates. If only a single bit is used for version tagging, then it is not possible to initiate a new update while an update is in progress. By using $2k$ version tag values it is possible to apply~$k$ separate updates concurrently under conventional version tagging schemes. The controller would then still need  to perform careful bookkeeping of the available version tag values, and after each update the controller would have to wait a sufficient period of time until all the `old' packets have been flushed from the network before it can reuse the old value. Using DPT all of this overhead can be avoided. 

DPT eliminates the need for the SDN application to manage per-flow or per-update version values. Instead, the time-of-day provides global versioning that is guaranteed to be monotonically non-decreasing, while allowing independence between updates.  

\subsection{Timestamp-based Load Balancing}
\label{LoadSec}
\begin{sloppypar}
The traditional Equal-Cost Multipath (ECMP) scheme, which balances traffic based on a hash of the packet header, has been in deployment for many years. This method has been shown to be inadequate for elephant flows~\cite{casado2013mice}.  Sophisticated methods such as~\cite{alizadeh2014conga} use performance monitoring to dynamically choose the least congested path. Random Packet Spraying (RPS)~\cite{dixit2013impact} has been suggested as a simple and stateless alternative for handling elephant flows.
\end{sloppypar}

DPT provides a simple mechanism for time-division-based traffic balancing, using periodic time ranges; since the timestamp is embedded in the packet header, we propose to use it as a forwarding criterion. Our approach is simple and stateless, and allows higher performance than existing stateless approaches, as shown in Sec.~\ref{EvalSec}.  

A simple example is illustrated in Fig.~\ref{fig:TwoLoadBal}a. Traffic from $S_1$ to $S_2$ is forwarded via three paths, A, B, and C. The capacity of each path is depicted in Fig.~\ref{fig:TwoLoadBal}a. 

\begin{figure}[htbp]
  \begin{subfigure}[t]{.23\textwidth}
	\centering
  \fbox{\includegraphics[height=3.8\grafflecm]{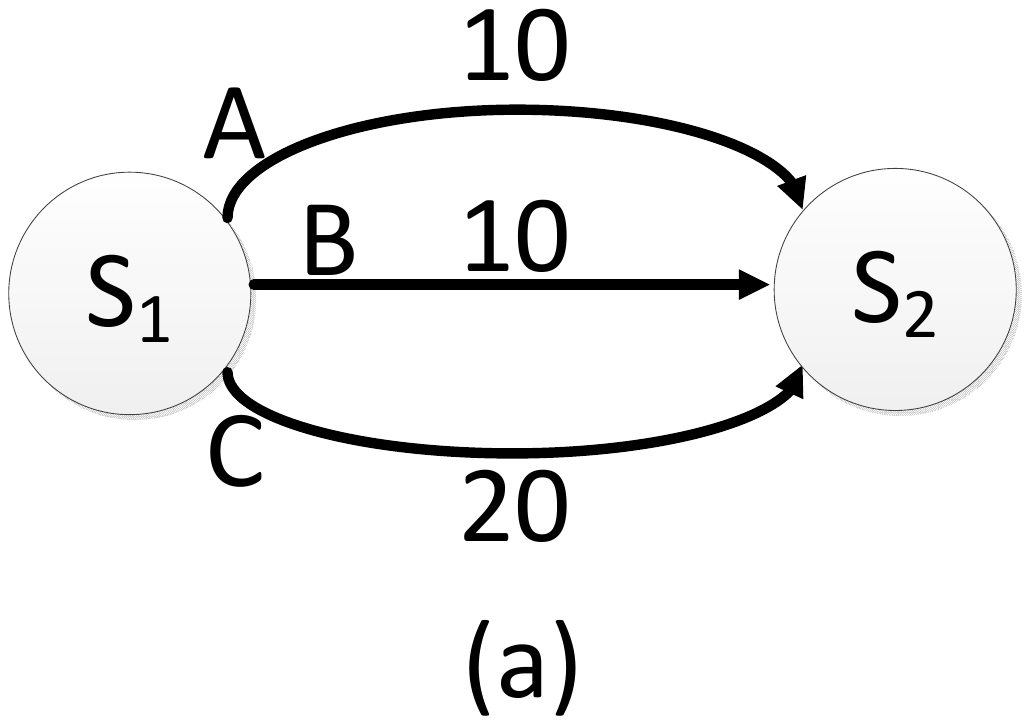}}
	\end{subfigure}%
  \begin{subfigure}[t]{.25\textwidth}
	\centering
  \fbox{\includegraphics[height=3.8\grafflecm]{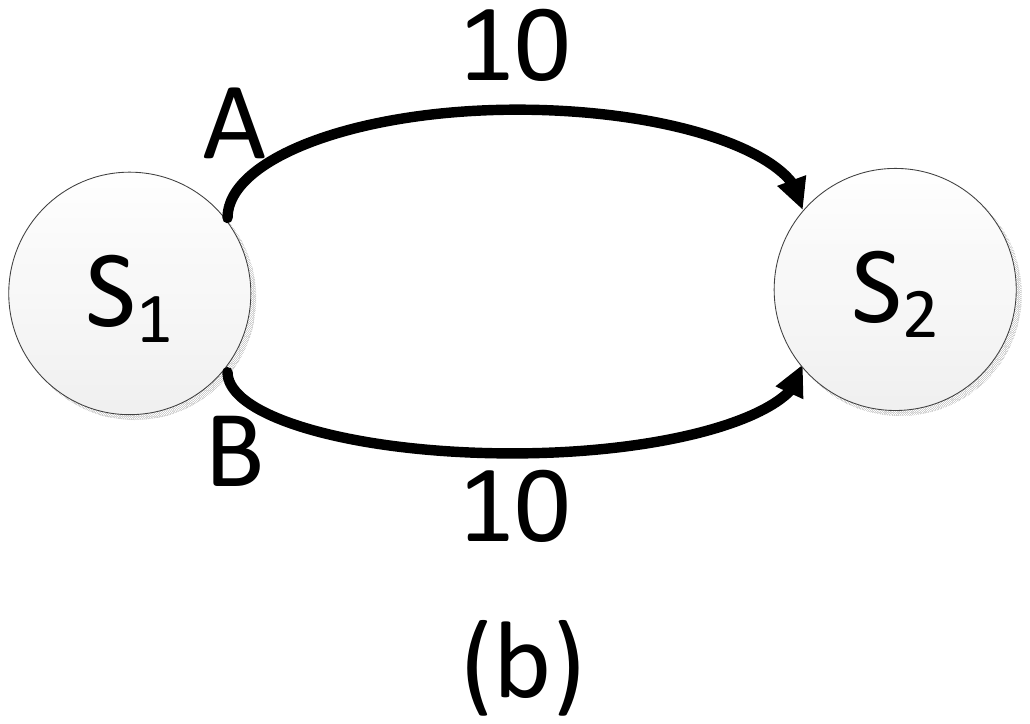}}
	\end{subfigure}
  \caption{Load balancing examples.}
  \label{fig:TwoLoadBal}
\end{figure}

The DPTH can be used for balancing traffic across the three paths; using periodic time ranges, traffic can be split over the three paths as illustrated in a time-division manner (Fig.~\ref{fig:BalPeriodic}), providing the desired weight to each path. 

\begin{figure}[htbp]
	\centering
  \fbox{\includegraphics[width=.45\textwidth]{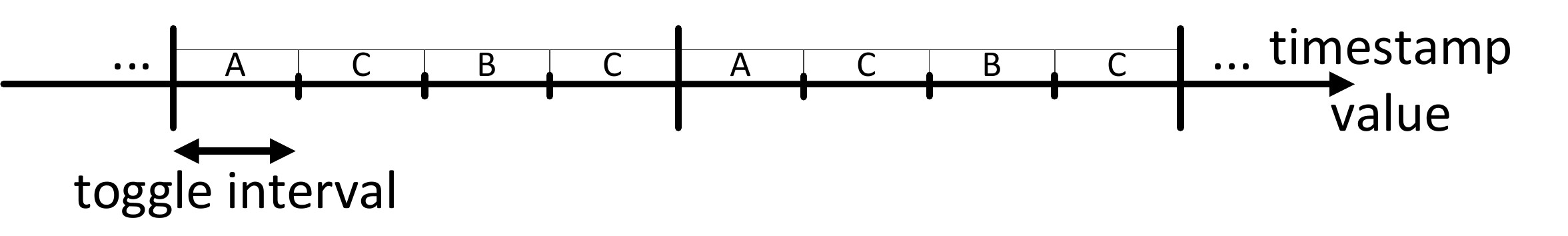}}
  \caption{Load balancing based on periodic timestamp ranges.}
  \label{fig:BalPeriodic}
	\ifdefined\CutSpace\vspace{-5mm}\fi
\end{figure}

\section{The Power of One Bit}
\label{OneBitSec}
In the previous section we presented three applications that can benefit from DPT. One of the main concerns that may arise from using time ranges is its \emph{cost} in terms of the number of timestamp bits added to each packet, and the number of match entries required to represent each time range. 

In this section we explore what can be done with a \textbf{single-bit} timestamp. 

\subsection{Network Telemetry using One Bit}
\label{Telemetry1BitSec}
In this subsection we assume that all data packets are timestamped with a 1-bit timestamp, which is the least significant bit of the Time.Sec field (Sec.~\ref{FormatSec}). Now the timestamp bit can simply be used as the color indicator, with a toggle interval of one second.

Each of the switches that take part in the measurement uses two separate match entries:

\begin{enumerate}
	\setcounter{enumi}{-1}
	\item Match: timestamp=0
	\item Match: timestamp=1
\end{enumerate}

Each entry has its own match counter, yielding a counter for color `0', and a counter for color `1'. The controller reads the counters once per second, allowing to compute the loss rate.

In this example the DPTH provides an inherent color bit without the need for the controller to be involved in the periodic color toggling. Moreover, each switch uses \textbf{a single} match rule per color.

\subsection{Consistent Updates using One Bit}
\label{Consistent1BitSec}
Previous work on dynamic traffic engineering~\cite{jin2014dynamic,jain2013b4} suggested that network paths should be reconfigured periodically, with a period on the order of a few minutes. As an example, we consider a 1-bit timestamp that represents the $7^{th}$ bit of the Time.Sec field, which toggles every 64 seconds. 

We propose to use this 1-bit timestamp as a version tag with a periodically toggled value, allowing an SDN traffic engineering application to apply a new set of network paths every 64 seconds. 

\begin{figure}[htbp]
	\centering
  \fbox{\includegraphics[width=.45\textwidth]{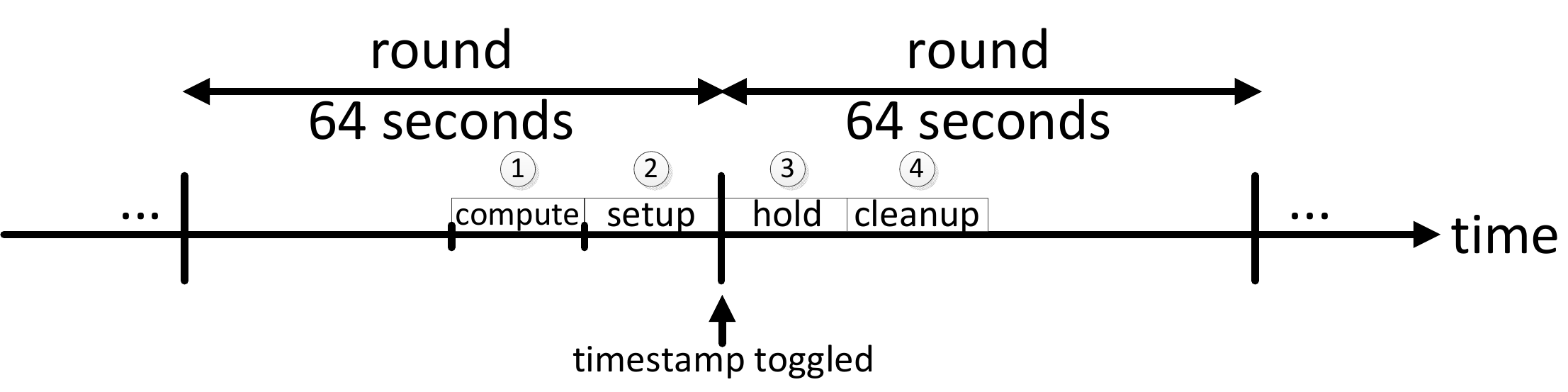}}
  \caption{Periodic consistent updates using a 1-bit timestamp.}
  \label{fig:PeriodicSche}
\end{figure}

\begin{figure*}[htbp]
	\centering
  \begin{subfigure}[t]{.33\textwidth}
  \centering
  \fbox{\includegraphics[height=6.25\grafflecm]{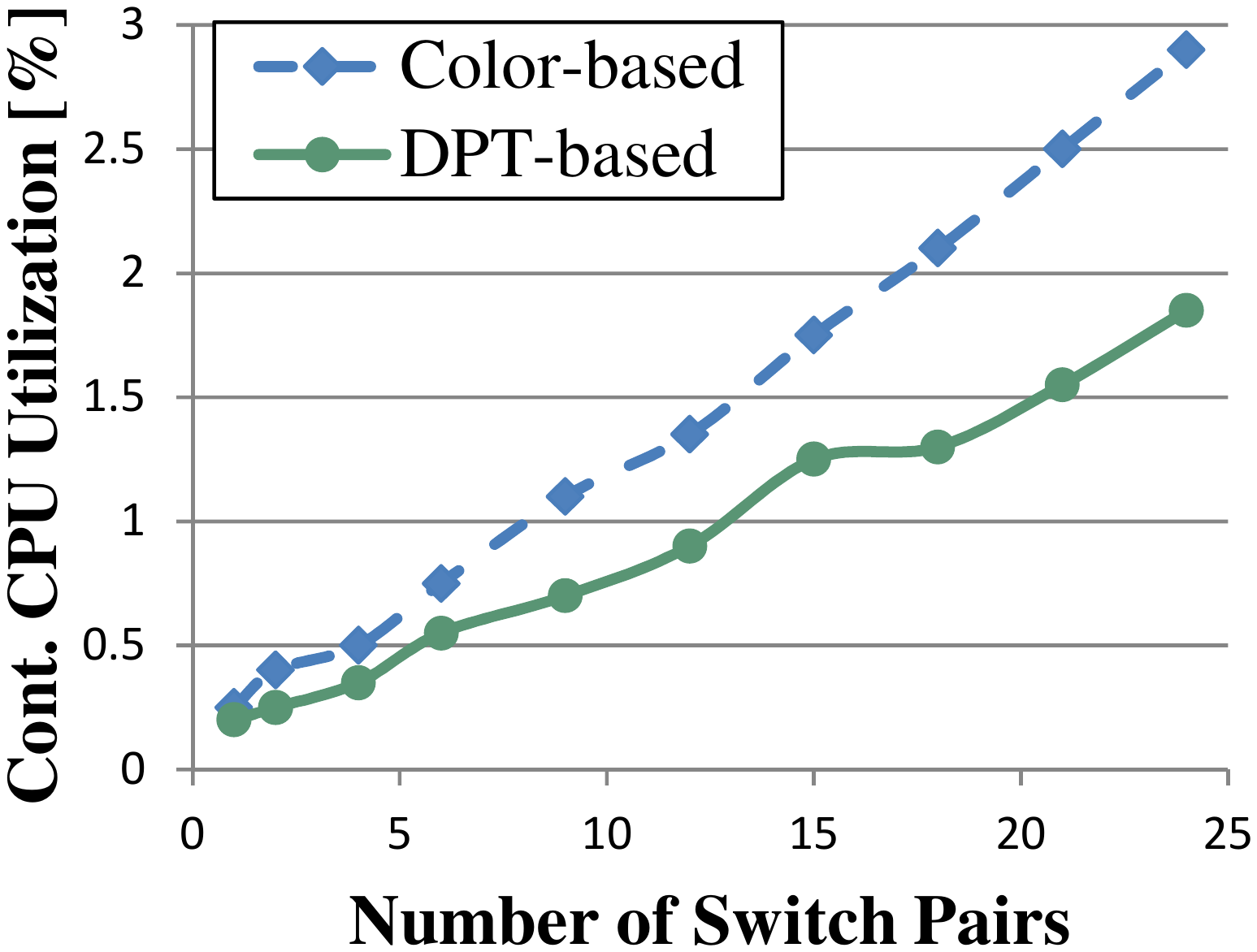}}
	\captionsetup{justification=centering}
  \caption{Experiment 1: telemetry.}
  \label{fig:TelemetryG}
  \end{subfigure}%
  \begin{subfigure}[t]{.33\textwidth}
  \centering
  \fbox{\includegraphics[height=6.25\grafflecm]{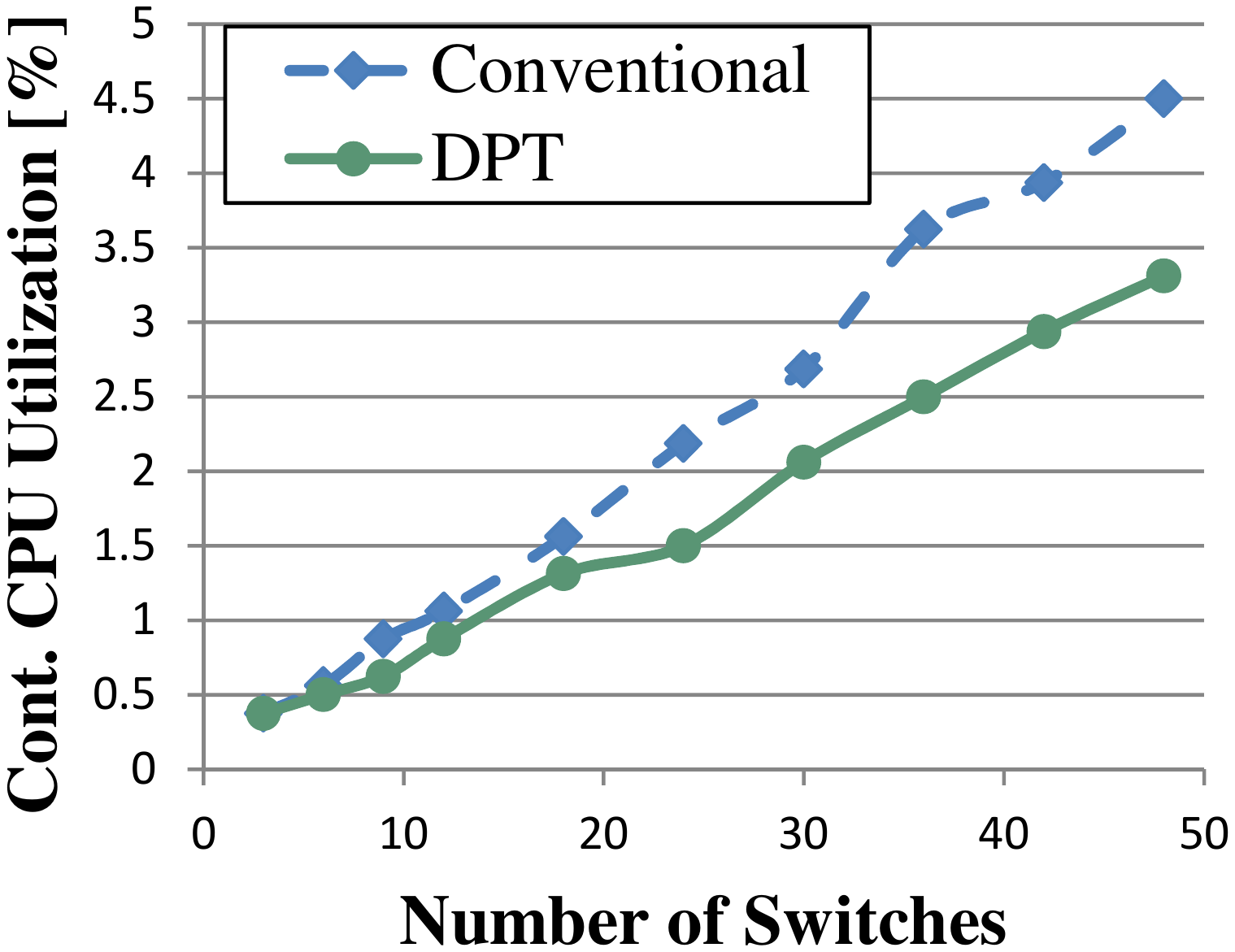}}
  \caption{Experiment 2: consistent updates.}
  \label{fig:UpdateG}
  \end{subfigure}%
  \begin{subfigure}[t]{.33\textwidth}
  \centering
  \fbox{\includegraphics[height=6.25\grafflecm]{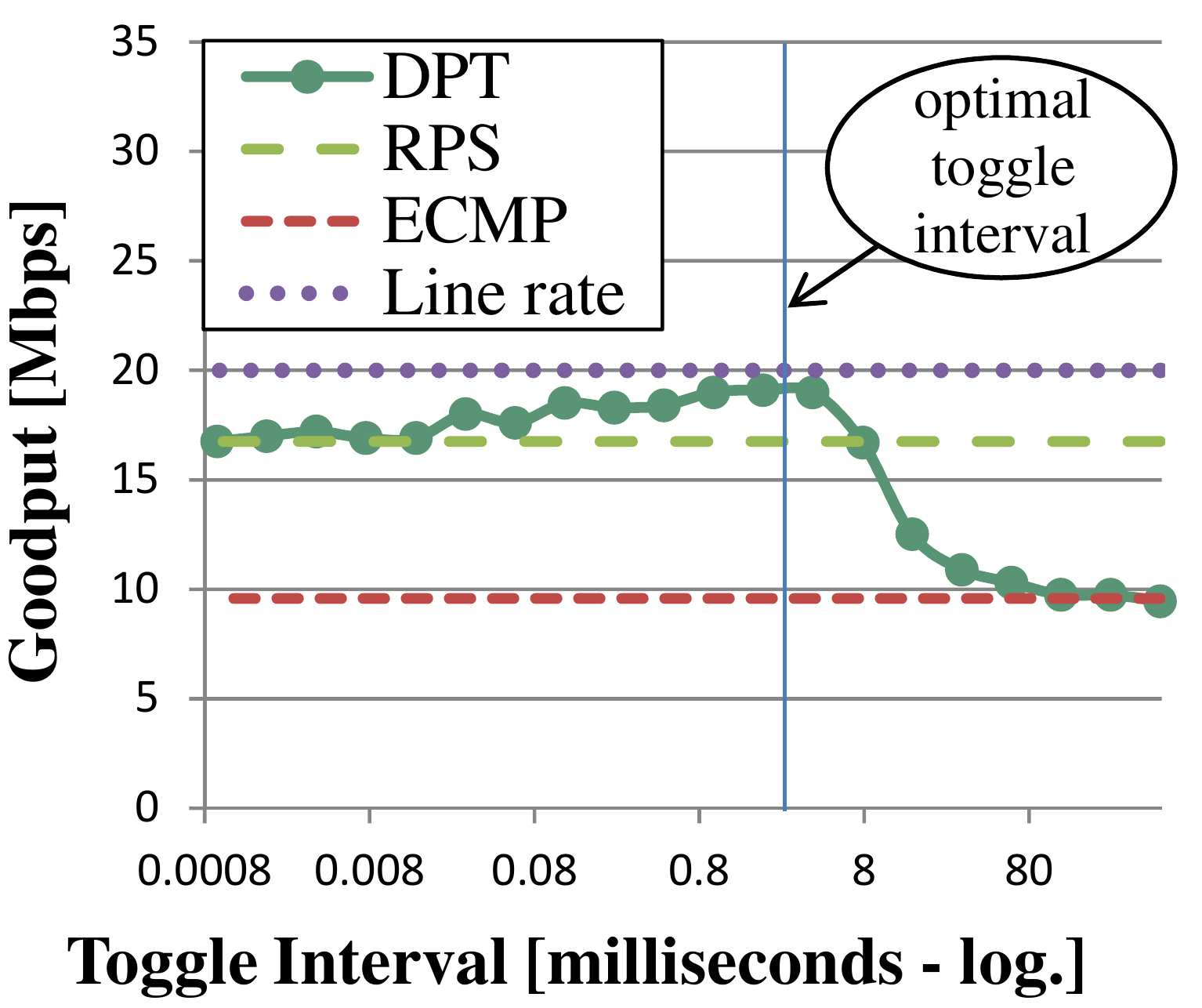}}
	\captionsetup{justification=centering}
  \caption{Experiment 3: load balancing. \\ Toggle interval is \textbf{chosen} by the SDN controller based on the link bandwidth, independent of the flow bandwidth.}
  \label{fig:LoadG}
  \end{subfigure}%
	\ifdefined\CutSpace\vspace{-5mm}\fi
  \caption{Experimental results.}
	\ifdefined\CutSpace\vspace{-5mm}\fi
  \label{fig:Graphs}
\end{figure*}

Fig.~\ref{fig:PeriodicSche} illustrates an example of the periodic schedule of an SDN application that uses this 1-bit timestamp. Each 64-second round is divided into four equal-sized slots: (1) the SDN application re-computes the network paths for the next round, (2) the new configuration is distributed to the switches and installed, (3) after the timestamp bit is toggled, the SDN controller waits until all the old packets have been drained from the network, and (4) the SDN controller removes the configuration of the previous round from the switches.

Each path that is modified between the two rounds requires two match entries during steps (2) and (3), one for the old configuration, and one for the new.

Since the DPTH plays the role of the version tag, the controller does not need to perform the second phase of the two-phase update (Fig.~\ref{fig:TwoPhaseAlgo}), thereby simplifying the `setup' process compared to conventional two-phase updates.

\subsection{Load Balancing using One Bit}
\label{LoadOneSec}
Consider the network of Fig.~\ref{fig:TwoLoadBal}b. An elephant flow of 20 Gbps needs to be balanced between two 10 Gbps paths. The controller can use timestamp-based rules that define a time division between the two paths. 

The total capacity of paths A and B is 20 Gbps. Assuming that packets are typically 1500 bytes long,\footnote{We assume that high-throughput flows are transmitted using maximal-length packets. In this example we assume that the Maximum Transmission Unit (MTU) is 1500 bytes.} we roughly have a packet every \textbf{0.6 microsecond}. We define a 1-bit timestamp that is the $12^{th}$ bit of the Time.Frac field. The toggle interval of this bit is roughly \textbf{0.5 microseconds}. The match rules in switch $S_1$ are defined to be:

\begin{enumerate}
	\setcounter{enumi}{-1}
	\item Match: timestamp=0. Action: output=path A.
	\item Match: timestamp=1. Action: output=path B.
\end{enumerate}

Intuitively, we would like the \textbf{toggle interval} to be roughly the same as the \textbf{inter-packet arrival interval}, as this allows packets to be forwarded in a balanced round-robin-like fashion, without the need for the stateful behavior of round-robin. Hence, configuring the correct toggle interval requires the controller to know the \textbf{total bandwidth} of the two paths, and the \textbf{MTU}.

\section{Evaluation}
\label{EvalSec}
In this section we present experimental evaluation of the three use cases that were presented in previous sections. The goal of these experiments is to quantify the advantages of using DPT in each of the use cases.

The experiments were performed on a testbed of 50 Linux-based machines in the Emulab~\cite{EmulabProj} environment. The experiments included 48 software-based OpenFlow switches, running the open source OFSoftSwitch~\cite{CPqDOF}, and Dpctl~\cite{Dpctl} was used as the controller. We slightly modified the code of OFSoftSwitch, allowing switches to embed the six least significant bytes of the NTP timestamp into the Ethernet source address (eth-src) of each packet. Since the timestamp was piggybacked onto the eth-src field, switches were able to perform match decisions based on this field.

\subsection{Experiment 1: Telemetry}
In this experiment we compared conventional color-based loss measurement~\cite{coloring} to the DPTH-based method of Sec.~\ref{Telemetry1BitSec}.

For each of the two methods we measured the controller's CPU utilization during a measurement of $M$ switch pairs, where each pair of switches performs the measurement as described in Sec.~\ref{Telemetry1BitSec}. We used the Linux `Top' utility to monitor the average CPU utilization. We used a {1-second} measurement interval, and ran each experiment for 100 seconds. We repeated the experiment for various values of~$M$.   

As shown in Fig.~\ref{fig:TelemetryG}, the controller's CPU utilization\footnote{The controller reaches a relatively high CPU utilization, around 3\% in an experiment with 48 switches, since the controller we used, Dpctl, is optimized for simplicity at the cost of performance.} when using DPT was approximately 30\% lower than in the color-based method. The difference is due to the fact that in the color-based method the controller needed to periodically (once per second) read the counters of each of the switches, and \textbf{also} to periodically modify the color of each monitored flow. In contrast, in the DPT-based method the controller does not need to modify the colors of each flow, as each of the switches derives the color from the DPT.

\subsection{Experiment 2: Consistent Updates}
We compare the controller's load in version-tag-based consistent updates~\cite{reitblatt2012abstractions} to the DPT-based updates of Sec.~\ref{Consistent1BitSec}.

We measured the CPU utilization during a periodic consistent update, as described in Sec.~\ref{Consistent1BitSec}, assuming an update period of 1~second.\footnote{In Sec.~\ref{Consistent1BitSec} we discussed an update period of 64~seconds. In this experiment we used a lower update period of 1~second in order to challenge the controller's performance.}

In each round (1~second), we performed a consistent update that involved $N$ switches. Each update included the following steps: (i) installing the new configuration in the $N$ switches, (ii) enabling the new version tag in the ingress switches (this step is only required in the non-DPT method), and (iii) removing the old configuration from the $N$ switches. We repeated the experiment for various values of $N$, up to~48, and each experiment was performed 100 times. As in~\cite{TimedConsistent}, we assumed that $\nicefrac{2}{3}$ of the switches are ingress switches, an assumption that is applicable to leaf-spine topologies such as Fat-Tree and Clos.
Fig.~\ref{fig:UpdateG} depicts the CPU utilization in each of the two methods. 

Notably, the DPT-based approach reduces the load on the controller's CPU by approximately 20\% compared to conventional two-phase updates, as the DPT method does not require step (ii) of the procedure above.

\subsection{Experiment 3: Load Balancing}
We used the topology of Fig.~\ref{fig:LoadBalExperiment}, and ran a TCP `elephant flow' of 20 Mbps from \emph{src} to \emph{dst} using Iperf~\cite{Iperf}. The capacity of each link (in Mbps) is depicted in the figure.

\begin{figure}[htbp]
	\centering
  \fbox{\includegraphics[width=.4\textwidth]{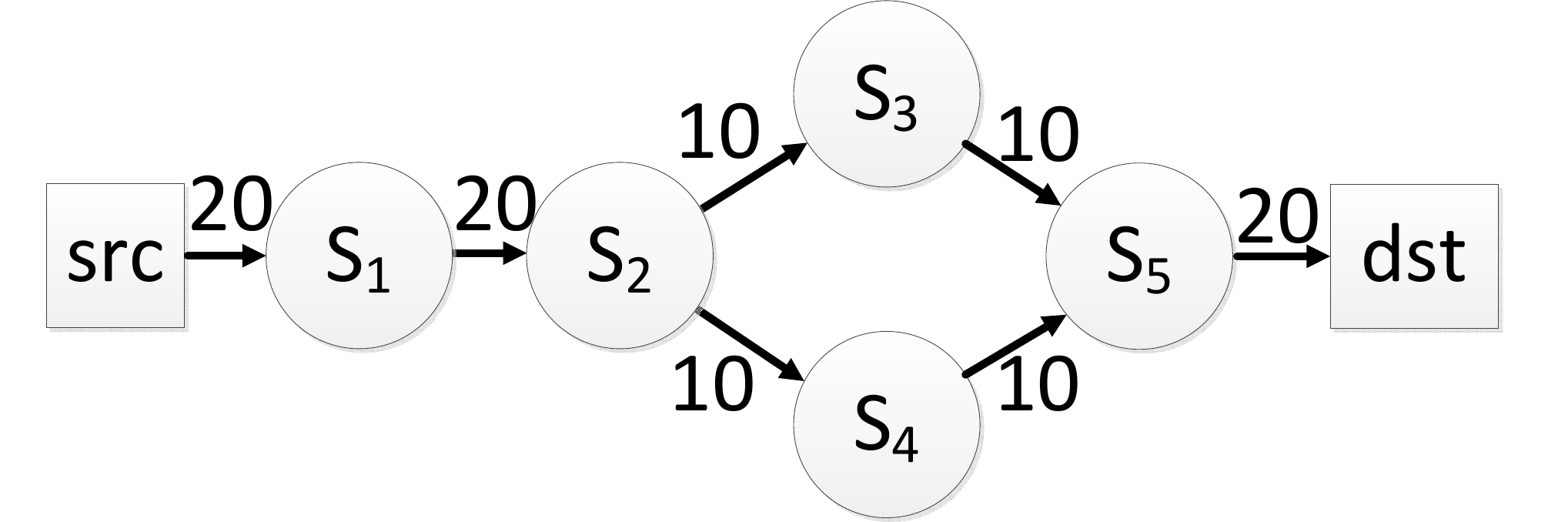}}
	\ifdefined\CutSpace\vspace{-1mm}\fi
  \caption{Load balancing experiment.}
  \label{fig:LoadBalExperiment}
	\ifdefined\CutSpace\vspace{-3mm}\fi
\end{figure}

Switch $S_1$ attached a timestamp to each packet, and $S_2$ performed the balancing between the two 10 Mbps links. In $S_2$ we used one-bit match rules, as described in Sec.~\ref{LoadOneSec}.
We repeated the experiment, each time matching a different bit in the timestamp field, causing a different \textbf{toggle interval}. In each experiment the traffic was run for a period of 10 seconds.

Fig.~\ref{fig:LoadG} illustrates the measured goodput of the TCP elephant flow as a function of the toggle interval. The figure shows the performance of the DPT-based method, compared to the two stateless approaches discussed in Sec.~\ref{LoadSec}: Random Packet Spraying (RPS)~\cite{dixit2013impact}, and conventional ECMP. 

Clearly, ECMP is not optimized for elephant flows, utilizing a bit under 50\% of the network capacity. RPS performs better than ECMP, but does not reach the full capacity of the links; since packets are sprayed randomly across the paths, each path is subjected to an occasional burst of consecutive packets that were incidentally forwarded to the same path, resulting in packet drops. These occasional packet drops cause the TCP algorithm to reduce the traffic rate.

\begin{sloppypar}
As shown in Fig.~\ref{fig:LoadG}, DPT performs significantly better than the other methods when the controller configures the toggle interval as discussed in Sec.~\ref{LoadOneSec}, i.e., when the toggle interval is roughly the same as the inter-packet interval.\footnote{The controller can compute the minimal inter-packet interval based on the \textbf{link} bandwidth, independent of the \textbf{flow} bandwidth, which varies over time.}
Decreasing the toggle interval gracefully reduces the performance, asymptotically resembling the performance of RPS. The reason is that using one of the lower bits of the timestamp is similar to using a randomized bit. On the other hand, as the toggle interval is increased beyond a few milliseconds the performance drops, since a large burst of packets is sent through a single path in each toggle interval, asymptotically behaving in an ECMP-like manner.
\end{sloppypar}

\ifdefined\TechReport
\section{Discussion}
\label{DiscussSec}
\subsection{Timestamp Size}
We showed that various applications can benefit from DPT, but what is the desired size of the timestamp field? Using a long timestamp field yields costly performance overhead. Sec.~\ref{OneBitSec} showed that a one-bit timestamp can suffice for some applications, but that each application may require the timestamp field to represent a different time scale. The size of the timestamp field presents a tradeoff: the most flexible solution, and also the most costly one, is to use a wide timestamp field, such as the format discussed in Sec.~\ref{FormatSec}. On the other hand, if the timestamp is known to be used for a specific application, a short timestamp field can provide the same functionality and reduce the on-the-wire overhead.

\subsection{Timestamps vs. Sequence Numbers}
\begin{sloppypar}
The approach we present suggests to include the time-of-day in every packet. Lamport~\cite{lamport1978time} suggested that distributed applications can use \emph{logical clocks}, which produce monotonically increasing \emph{sequence numbers} instead of \emph{timestamps}. 
\end{sloppypar}

\textbf{The advantage of timestamps.} 
We argue that there are advantages to using a DPT that represents the time-of-day. For instance, in the telemetry use case the DPT defines the traffic blocks to be at fixed time intervals, allowing the controller to sample the counter and timestamp values at fixed time intervals. In the consistent update use case the use of time-of-day allows the controller to plan a schedule of when the update rules need to be installed and removed.

\begin{sloppypar}
\textbf{Clock accuracy.} Using timestamps implies that switches maintain accurately synchronized clocks. Notably, if clocks are not synchronized, then using timestamps is equivalent to using sequence numbers. The required accuracy of the clocks depends on the application that uses DPT. For example, the telemetry application of Sec.~\ref{Telemetry1BitSec} requires an accuracy that is better than 1~second, as the controller retrieves the measurement information from the switches once per second. The consistent update application described in Sec.~\ref{Consistent1BitSec} will work well even if the clock accuracy is on the order of several seconds.
\end{sloppypar}

\subsection{Impact on Network Protocols}
As mentioned in Sec.~\ref{UsingTimestampsSec}, the evolving P4 language is already designed in a DPT-friendly way. It is also important to consider how DPT affects the on-the-wire protocols.

\textbf{Data plane protocol.} Since the DPTH is designed to be added to every packet, it must be supported by the data plane encapsulation. Fortunately, many of the recently defined data plane encapsulation protocols~\cite{vxlan,geneve,nsh} have flexible support for type-length-value (TLV) or metadata fields in the packet header. Therefore, it is relatively straightforward to define such TLV or metadata fields for the DPT in each of these encapsulations.

\textbf{Control plane protocol.} In order to support DPT, two main extensions are required in OpenFlow~\cite{OpenFlow1.5}. Based on the current activity in the Open Networking Foundation (ONF) and in the P4 consortium, we believe that both features are imminent in future versions of OpenFlow.
\ifdefined\TechReport
\begin{itemize}
\else
\begin{itemize}[leftmargin=*]
\fi
	\item The ability to add or remove a timestamp from the packet header. The ability to store a packet's time of transmission or reception is also required, as described in Sec.~\ref{TelemetrySec}. We note that the ONF is currently working on OAM support~\cite{ONFCarrier}, which requires similar timestamping capabilities, and thus these capabilities are likely to be added in future versions of OpenFlow.
	\item The ability to use the DPTH in match procedures. The ONF's current work on Protocol Independent Forwarding (PIF)~\cite{ONFPIF} will allow the OpenFlow match procedure to be based on any field in the packet header, including the DPTH.
\end{itemize}

\subsection{Security Considerations}
The security considerations of using DPT are discussed in detail in~\cite{nshtimestamp}. In-band timestamping can be used as a means for network reconnaissance. By passively eavesdropping to timestamped traffic, an attacker can gather information about network delays and performance bottlenecks.

The DPT is intended to be used by various diverse applications. Thus, a man-in-the-middle attacker can maliciously modify timestamps in order to attack applications that use the timestamp values. 

DPT relies on an underlying time synchronization protocol. Thus, if the time protocol is not properly secured~\cite{mizrahi2013security}, then by attacking the time protocol an attack can potentially compromise the integrity of the DPT. 

\fi

\section{Conclusion}
DPT provides a single header field that can be used for multiple purposes. 
We have shown three interesting use cases for DPT, and we believe that DPT can be useful for various other SDN applications.
Although adding a DPT header to all packets appears to be costly at a first glance, we have shown that the cost of DPT is often reduced to a single bit. Our work suggests that the benefits of DPT in SDNs certainly outweigh the costs. 

\textbf{Acknowledgments.}
The authors gratefully acknowledge Danny Raz and Boaz Patt-Shamir for useful discussions.
We thank the Emulab project~\cite{EmulabProj} for the opportunity to perform our experiments on the Emulab testbed. This work was supported in part by the ISF grant 1520/11.

\bibliographystyle{ieeetr}
\ifdefined\TechReport
\bibliography{TimestampAll}
\else
\bibliography{TimestampAllShort}
\fi

\end{document}